\documentclass{article}
\usepackage{lnfprep}
\usepackage{graphicx}
\usepackage{subfigure}
\usepackage{amsmath}
\usepackage{amssymb}

\def\beq{\begin{equation}}   \def\eeq{\end{equation}}
\def\bea{\begin{eqnarray}}   \def\eea{\end{eqnarray}}

\newcommand{\gsim}{\lower.7ex\hbox{$ \;\stackrel{\textstyle>}{\sim}\;$}}
\newcommand{\lsim}{\lower.7ex\hbox{$ \;\stackrel{\textstyle<}{\sim}\;$}}

\def\c2{CLEO~II.V}

\def\d0d0{ D^0\bar{D}^0 }
\def\p0p0{ P^0\bar{P}^0 }

\def\qp2{ \Bigl| \frac{q}{p} \Bigr|^2 }
\def\pq2{ \Bigl| \frac{p}{q} \Bigr|^2 }

\def\ps2s{  \psi(2S) }
\def\q2{ $q^2$ }
\def\cm2s1{ $\,{\rm cm}^{-2} {\rm s}^{-1}$}
\def\d0{D_2^{*0}}
\def\d+{D_2^{*+}}

\newcommand{\Header}{
  \begin{tabular}{rl}
  \hspace{-.4cm}
      &
    \renewcommand{\arraystretch}{0.5}
    \renewcommand{\arraystretch}{1}
  \end{tabular}
  \vskip 1cm
  \begin{flushright}
  \renewcommand{\arraystretch}{0.5}
    \begin{tabular}{r}
      {\underline{LNF-05/32 (P)}}\\    
      {\small 27 dicembre 2005} \\      
    \end{tabular}
  \end{flushright}
  \renewcommand{\arraystretch}{1}
  \vskip 1 cm
  }

\begin{document}
\begin{titlepage}
\title{
  \Header
  {\LARGE  \textsc{\textmd {A Novel Approach for an Integrated Straw
  tube-Microstrip Detector}} 
  }
}
\author{E.Basile(*), F.Bellucci (***), L.Benussi, M. Bertani, S. Bianco, M.A.
Caponero (**), D. Colonna (*), \\ 
F. Di Falco (*), F.L. Fabbri, F. Felli (*), M.Giardoni, A. La Monaca, G.
Mensitieri (***), B. Ortenzi, \\ 
M. Pallotta, A. Paolozzi (*), L.Passamonti, D.Pierluigi, C. Pucci (*), A. Russo,
G. Saviano (*), F. Massa(****).\\ 
{\em Laboratori Nazionali di Frascati dell'INFN }\\
\\
\\
F.Casali, M.Bettuzzi, D.Bianconi
\\
{\em University of Bologna and INFN, Bologna, Italy }\\
\\
\\
F. Baruffaldi, E. Petrilli \\
{\em Laboratorio di Tecnologia Medica, Ist. Ortop. Rizzoli and
University of Bologna, Bologna, Italy }}
 \maketitle
\baselineskip=1pt
\begin{abstract}
\indent \indent We report on a novel concept of silicon microstrips
and straw tubes detector, where integration is accomplished by a
straw module with straws not subjected to mechanical tension in a
Rohacell $^{ \circledR}$ lattice and carbon fiber reinforced plastic
shell. Results on mechanical and test beam performances are reported
on as well.
\end{abstract}

\vspace*{\stretch{2}}
\begin{flushleft}
  \vskip 2cm
{ Index Term: Elementary Particles, Detectors, Tracking.}
\end{flushleft}
\begin{center}
\emph{Submitted to Transactions on Nuclear Science}
\end{center}

\vskip 1cm
\begin{flushleft}
\begin{tabular}{l l}
  \hline
  $ ^{*\,\,\,\,\,\,}$ & \footnotesize{Permanent address: ``La Sapienza" University - Rome.} \\
  $ ^{**\,\,\,}$& \footnotesize{Permanent address: ENEA Frascati.} \\
  $ ^{***}$ & \footnotesize{Permanent address: ``Federico II" University - Naples.} \\
  $ ^{****}$ & \footnotesize{Permanent address: INFN Rome 1.} \\
\end{tabular}
\end{flushleft}
\end{titlepage}
\pagestyle{plain}
\setcounter{page}2
\baselineskip=17pt

\section{  \textsc{Concept}}
Modern physics detectors are based on tracking subcomponents, such
as silicon pixels and strips, straw tubes and drift chambers, which
require high space resolution, large geometrical acceptance and
extremely large-scale integration. Detectors are often requested
with demanding requirements of hermeticity and compactness that must
satisfy the minimization of materials. We have developed for the
BTeV experiment[1] an integrated solution that accommodates straw
tubes and silicon strips in a common structure. Our novel design
utilizes glued straw tubes mechanically un-tensioned and embedded in
a Rohacell$^{ \circledR}$ lattice. The straw-Rohacell$^{ \circledR}$
composite is enclosed in a carbon fiber reinforced plastic (CFRP)
shell and supports the microstrip detector. Un-tensioned straw tubes
have been used in the past [2]. The untensioned straws in the ATLAS
TRT detector [3] at LHC need support dividers every 25cm of their
lengths, and  four sets of thin  carbon fiber filaments each to
provide required stiffness. The TOF detector [4] at COSY uses a very
large gas overpressure (2bar) for straw stiffness. Our design avoids
supports and filaments, operates straws in standard conditions of
very small gas overpressure, and allows integration between straws
and microstrip
detectors.\\
\newline
\section{\textsc{BTeV Detector}}
The BTeV experiment[1] at the Fermilab proton-antiproton collider
(the Tevatron) will produce and study particles containing the
beauty heavy quark, in order to investigate the phenomenon called CP
violation, and understand if the Standard Model of particles and
interactions is sufficient to describe the world we live in. BTeV is
composed of tracking detectors  (pixel, strips, straws) for
detection of charged particles, a RICH Cerenkov detector for
identification of pions, kaons and protons, a crystal EM calorimeter
for detection of neutral particles (photons, $\pi^0$) and
identification of electrons, and a muon detector. The experimental
setup is shown in Fig.1.
 \newline
\section{  \textsc{M0X Concept}}
The M0X module is a special module to be placed closest to the beam,
measuring the x-coordinate of tracks.  It houses straw tubes and
supports silicon strip detector planes. M0X is made of straw tubes
embedded in a Rohacell$^{ \circledR}$  IG50 foam, inside a CFRP
shell. Straws are glued together and onto the foam, the foam is
glued to the CFRP shell. CFRP is chosen to allow the fabrication of
a rigid mechanical structure with high transparency to incoming
particles. CFRP is also used for the M1 modules, conventional straw
tubes sub-detectors that act as struts sustaining the mechanical
tension of the remaining straw modules. Six straw-microstrip
stations are deployed in BTeV, each station made of three views (X,
U, V), each view made of two half-views. The X view (vertical
straws) measures the X coordinate, while the two stereo views (U,V)
are at $\pm11.3^\circ$ around the Y bend coordinate. Straw inner
diameter is 4mm, straw lengths vary from 54cm in the first station
to 231cm in the sixth station. Fig.2 shows a conceptual  design of
M0X and M1X assembly (left), and a 60-cm-long prototype with straws
embedded in Rohacell$^{ \circledR}$ (right).
\newline
\section{  \textsc{Tomography}}
A check of the eccentricity of the straws and of their positions in
the grooves can be done with a tomography method. The tomography
uses X-rays and can reconstruct sections of the scanned region. The
technique determines location and geometry of straws by
reconstructing images of their cross sections. Computed images are
reconstructed from a large number of measurements of X-ray
transmission. Reconstruction provides 2-dimensional and
3-dimensional images of straws. 2-dimensional images of 6-channel
M0X prototype, 3-dimensional reconstruction, and 2-dimensional image
of final assembly technique M0X prototype are shown in Figure 3
left, center, right, respectively. Preliminary results show that a
precision of about $20\mu$m can be reached on the measurement of
straw radii. The maximum variation from circularity allowed is $100
\mu$m, in order not to change the straw gain by more than 10\%.
\newline
\section{  \textsc{Finite Element Analysis}}
A finite element analysis (FEA) [5],[6] allows  to estimate the
displacements of the M0X module under the loads of the microstrip
detectors and straw tubes. Time stability and maximum displacements
of the order of $10\,\mu$m are requested, in order not to spoil the space
resolution of the microstrip detectors. The FEA analysis has been
carried on the M0X of the sixth station, the longest straw length. A
straw load of 12N in each corner of M0X has been simulated to
reproduce the mechanical tension of wires. A straw load of 12N and a
torque of 2Nm have been applied to simulate the weight of the
micro-strip. The geometry and mechanical properties of materials
used are reported in Tab.1. Fig.4 shows the graphical output of FEA
with isocurves of deformation under simulated load of microstrip
detector and associated electronics. Maximum deformation is pointed
to by arrow in Fig.4. FEA shows a maximum displacement of about
$15\mu$m ($4\mu$m in the axial direction, $9\mu$m x direction,
$11\mu$m y direction), close to the required specification. We have
used shell elements for the simulation of the carbon fiber
reinforced polyester structure and bricks for the Rohacell$^{
\circledR}$ simulation. Beam elements were used for introducing glue
between the CRFP module and the cylindrical plate where microstrips
were placed. The geometry shown in Tab.1 and Fig.4 corresponds
 to a $0.007X_0$ material thickness in radiation lengths units, which represents a 60\%
 reduction in material with respect to a design with independent supports.
\newline
\begin{table}[!h]
\begin{center}
\begin{tabular}{|c||c|c|c|}
\hline

 & M\textsc{0 and M1 Structure} & \textsc{Micro Strip Cylinder}  & \textsc{Rohacell Foam} \\
\hline
\hline
\textsc{Thickness} & 0.07 \textsc{each ply with} & 0.07 \textsc{each ply} (0°/90°/0°) & \\
\textsc{[mm]} & \textsc{fibres disposition} & \textsc{with a rohacell}& 2 \\
  & \textsc{of} 0°/90°/0° & \textsc{foam of 5cm}&  \\
\hline
$E_{11}[GPa]$ & 260 & 590 & 0.019 \\
\hline
$E_{22}[GPa]$ & 10 & 10 & - \\
\hline
$E_{12}[GPa]$ & 7.2 & 7.2 & - \\
\hline
$\nu_{12}$ & 0.3 & 0.3 & 0.3 \\
\hline
\end{tabular}
\caption{Geometrical arrangement (thicknesses), Young modules
($E_{11}$, $E_{12}$, $E_{22}$), and Poisson coefficient $(\nu_{12})$
used in FEA simulation of M0X.}
\end{center}
\end{table}
\section{  \textsc{FBG Sensors}}
Fiber Bragg Grating (FBG) sensors have been used so far as
telecommunication filters, and as optical strain gauges in civil and
aerospace engineering [7], and, only recently, in HEP detectors [8].
The BTeV detectors utilize Fiber Bragg Grating (FBG) sensors to
monitor online the positions of the straws and microstrips. The
optical fiber is used for monitoring displacements and strains in
mechanical structures such as the straw tube-microstrip support
presented here. A modulated refractive index along the FBG sensor
produces Bragg reflection at a wavelength dependent on the strain in
the fiber, permitting real-time monitoring of the support. According
to these properties, an FBG sensor is going to be placed in the M0X
structure between the Rohacell$^{ \circledR}$ foam and the CFRP
shell. Sensors will be located  in spots of maximal deformation, as
predicted by FEA simulation. Fig.5 shows long-term behaviour of FBG
sensors while monitoring micron-size displacements, compared to
monitoring via photographic methods.
\newline
\section{  \textsc{Prototype}}
M0X prototypes have been fabricated in order to study the
construction procedures, mechanical properties, material
characterization, and physical behaviour for detection of particles
in test beam set-ups. Straw materials such as mylar and kapton have
been studied and characterized for tensile properties under exposure
to Ar-CO$_2$ mixtures [9]. The most demanding design requirement is
the assembly of straws in a close pack, with no mechanical tension
applied. Several gluing techniques have been examined and tested to
determine the optimal technique. Straw tubes are glued together in
three layers, and the upper and lower layer are glued to the
Rohacell$^{ \circledR}$ foam.  Glues with different viscosity, and
several gluing techniques, have been used.  Glues tested range from
cyanoacrylate (Loctite 401) to epoxy (Eccobond series). Gluing
techniques ranged from brush, to injection, to spray gluing. The
most promising results have been obtained by using an Eccobond 45W
and catalyst mixture (1:1 by weight), diluted with dimethylcheton
solvent. For each 40g of glue-catalyst mixture, 40cm$^3$ of solvent
was used.\\
The assembly process proceeds as follows. Stainless
steel rods (4-mm-diameter) are inserted in each straw tube. A straw
layer is formed by locating 16 straws on a machined grooved plate.
The glue-solvent mixture described is sprayed, with 2bar air
pressure, and 20cm distance between spray gun and straw layer.
Mechanical pressure is applied to layers during curing. After curing
at room temperature, the straw layers are sprayed again and more
layers are superimposed. After additional curing, the stainless
steel rods are removed from the straws. The mechanical pressure
applied during curing allows loose (0.1\%) requirements on rod
straightness. Conductive contact between straw cathodes and
aluminum endplate, is accomplished via spraying of Eccobond 57C.
Fig.6 shows a complete prototype after gluing, with endplate for
wiring on one side. The glued layers of straws provide excellent
stiffness (see Sect. Tomography) for operation, without need of
Rohacell$^{ \circledR}$ foam which, instead, contributes to
mechanical properties when used in integration with microstrip
detector.  Fig.7 shows detail of the CFRP shell near the beam pipe
region (left), and the grooved foam (right).
\newline
\section{\textsc{Cosmic Ray And Test Beam Results}}
Preliminary results with cosmic rays show very clean pulses (Fig.8)
in gas mixtures of interest for BTeV (Ar-CO$_2$ 80/20), with shape
parameters typical of operation with such mixtures. High-voltage
applied is +1400V, a trans-impedance preamplifier [10] provides a
2V/mA gain, followed by a low-walk double-threshold discriminator
[11]. Prototypes have been exposed to beam particles in the Frascati
Test Beam Facility [12]. Preliminary results show the expected
response of prototype to minimum ionizing particles. The
distribution of drift time of the ionization electrons to the sense
wire (Fig.9 left) over the straw 2mm radius is compatible with the
drift velocity in the Ar-CO$_2$ (80/20) gas mixture used. A 10mV
threshold is applied. Beam particles tracks are reconstructed by the
M0X prototype, tracks residuals are shown in Fig.9 right. The
distribution of residuals is well fitted by a Gaussian with a rms
width of about 130$\mu$m.
\newline
\section{  \textsc{Conclusions}}
We have developed a novel concept for integration of straw tubes
tracking detectors and silicon microstrip detectors, for use in HEP
experiments at hadron colliders. In our design, silicon microstrips
are integrated to a special  straw tube module M0X via a CFRP
mechanical structure. M0X is realized via glued straws embedded in a
Rohacell$^{ \circledR}$ lattice with no need of mechanical tension.
Detailed finite element analysis shows that deformations affect
negligibly the tracking performance of the system. A complete system
based on Fiber Bragg Grating sensors - acting as optical strain
gauges - monitors the position of each sub detector with  micron
resolution. Test beam studies are underway to verify that M0X can
provide the 200$\mu$m resolution needed by the BTeV tracking
detector requirements.
\newline
\section*{  \textsc{Acknowledgements}}
\indent We thank G.Mazzitelli (LNF INFN, Italy) and all the
DA$\Phi$NE team for smooth running of the Beam Test Facility.
\newline

\newpage
\begin{figure}[!htbp]
\begin{center}
  \includegraphics[width=10cm]{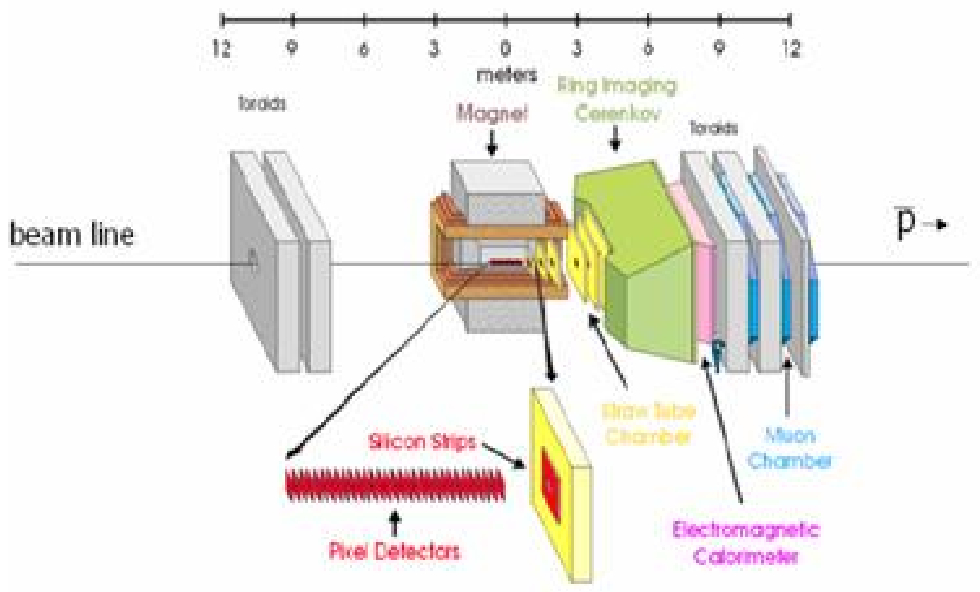}\\
  \caption{BTeV detector layout; the straw tube chamber are in yellow, the silicon strips in red.}
  \end{center}
\end{figure}

\

\begin{figure}[!htbp]
\begin{center}
  \includegraphics[width=15cm]{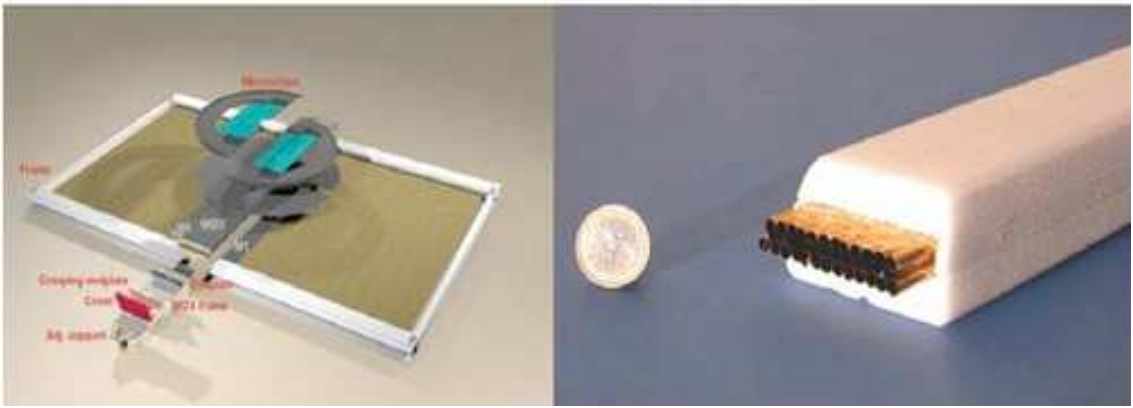}\\
  \caption{Exploded view of BTeV microstrip and straws tubes integration (left);
  M0X prototype with straw tubes embedded in  Rohacell $^{
\circledR}$ (right).}
  \end{center}
\end{figure}

\begin{figure}
\begin{center}
  \includegraphics[width=15cm]{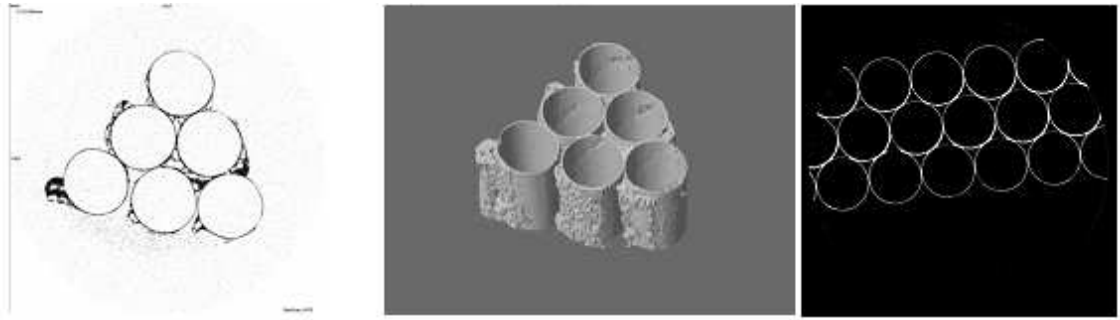}\\
  \caption{2-dimensional images of 6-channel M0X prototype (left);
  3-dimensional reconstruction (center); 2-dimensional image of final
  assembly technique M0X prototype (right). Tomographs are for straws
  at the foam-supported stage.}
  \end{center}
\end{figure}

\begin{figure}
\begin{center}
  \includegraphics[width=10cm]{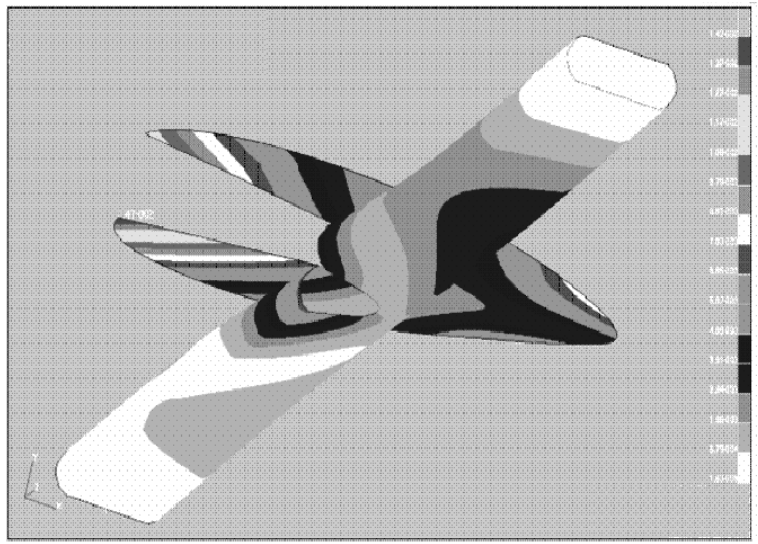}\\
  \caption{FEA results for the simulation of M0X (straws-Rohacell $^{
\circledR}$-CFRP) straw-microstrip detector. Geometry and materials
are shown in Tab.1.
  Colour levels show curves of  equal deformation under simulated load of
  microstrip detector and associated electronics.  Units are micrometers.
  Maximum deformation is 14.7 $\mu$m (arrow).}
  \end{center}
\end{figure}

\begin{figure}
\begin{center}
  \includegraphics[width=12cm]{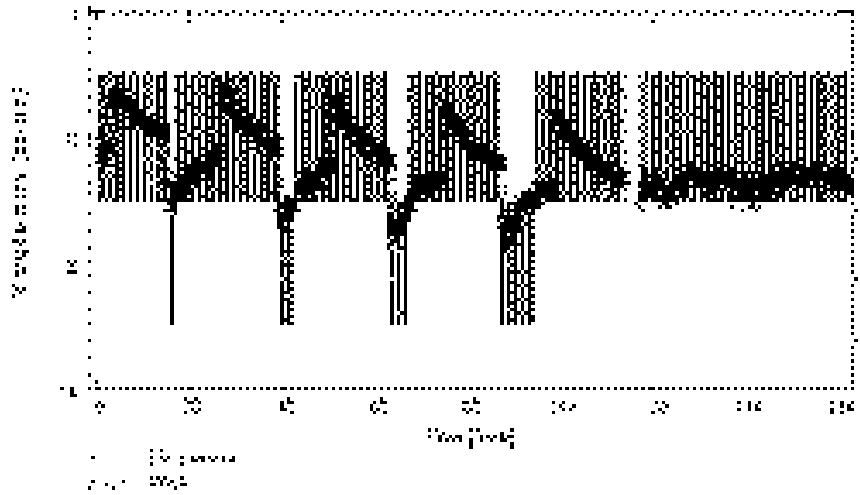}\\
  \caption{FBG long-term monitoring stability results.
  FBG output (crosses)  is validated by TV camera (bars).
  The  bar size indicates the best resolution of TV camera.}
  \end{center}
\end{figure}

\begin{figure}
\begin{center}
  \includegraphics[width=15cm]{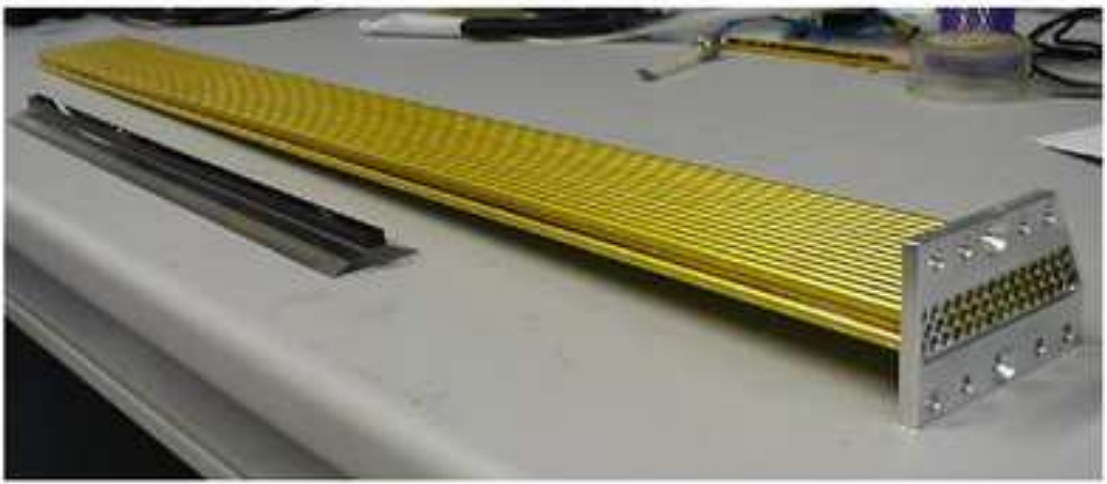}\\
  \caption{M0X module prototype during assembly.  Straw tubes
  are glued together and positioned between end-plates (one shown)
  without mechanical tension. Rohacell$^{
\circledR}$ foam and CFRP shell not shown.}
  \end{center}
\end{figure}

\begin{figure}
\begin{center}
  \includegraphics[width=15cm]{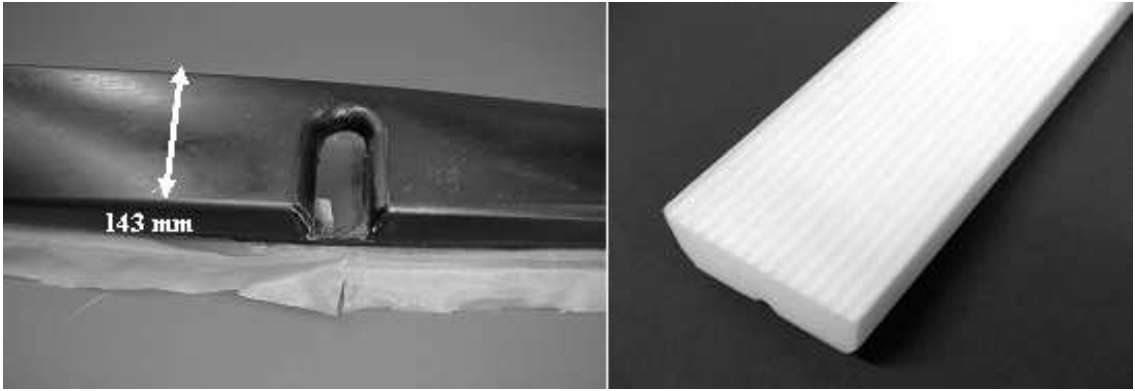}\\
  \caption{CFRP shell prototype immediately after fabrication (left); grooved Rohacell$^{
\circledR}$ foam (right).}
  \end{center}
\end{figure}

\begin{figure}
\begin{center}
  \includegraphics[width=8cm]{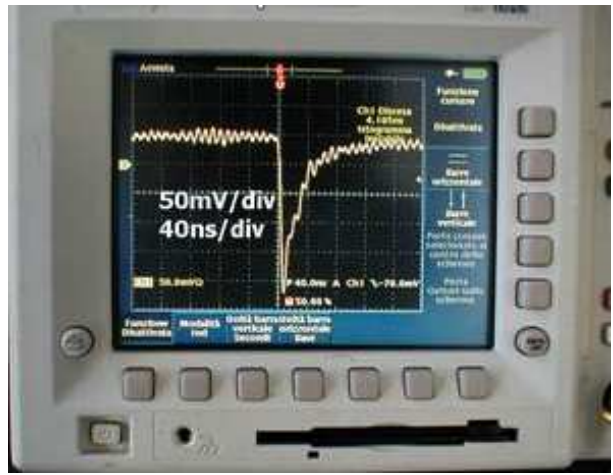}\\
  \caption{Cosmic rays signals in M0X prototype with
  (Ar-CO$_2$ 80/20) gas mixture. High-voltage applied is 1400V,
  a transimpedance preamplifier provides a $2V/mA$ gain.}
  \end{center}
\end{figure}
\begin{figure}
\begin{center}
  \includegraphics[width=15cm]{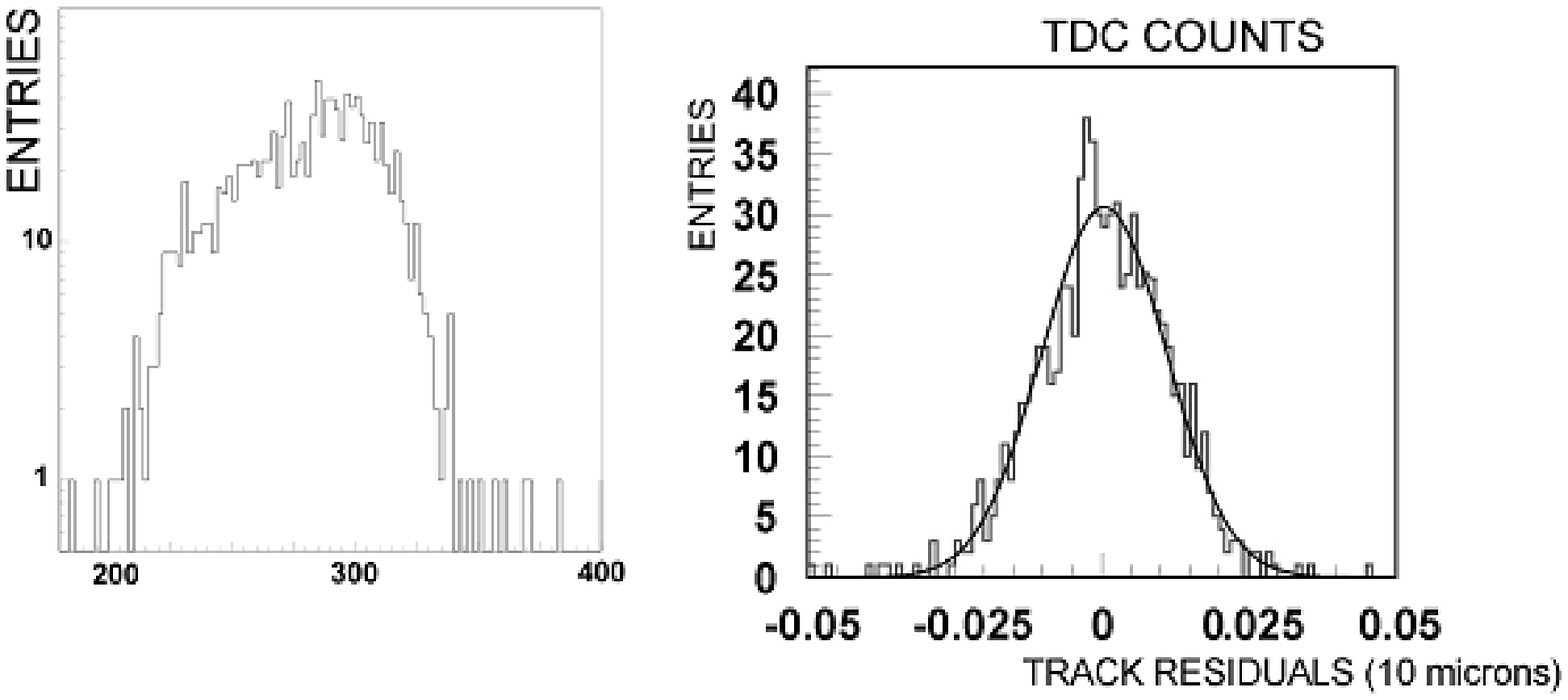}\\
  \caption{Distribution of drift times from beam particles.
  Times are expressed in time-to-digital-converter counts (300ps/count),  Ar-CO$_2$
  (80/20) gas mixture is used, with 10mV threshold (left);
  tracks residuals of beam particles tracks  reconstructed by the M0X prototype,
  the distribution of residuals is well fitted by a Gaussian with a rms width of about 130$\mu$m (right).}
  \end{center}
\end{figure}

\begin{thebibliography}{0}

\bibitem
.Fermilab Experiment E-0897/E-0918, J.Butler, S. Stone
co-spokespersonsw; see www-btev.fnal.com.
\bibitem
.S.H.Oh et al., Nucl. Instr. Meth. A309 (1991) 368-376.
\bibitem
.T. Akesson et al., Nucl. Instr. Meth. A522 (2004) 131-145.
\bibitem
.K. Nuenighoff et al., Nucl. Instr. Meth. A477 (2002) 410-413.
\bibitem
.E. Basile, "Scelta dei materiali ed analisi strutturale per
supporti di rivelatori di particelle dell'esperimento BTeV a
Fermilab (U.S.A.)", degree thesis, University "La Sapienza", Rome,
2003 (in Italian). Also available at
http://www-btev.fnal.gov/cgi-bin/public/DocDB/ShowDocument

\bibitem
.C. Pucci, "Analisi strutturale del supporto per microstrip straw
tubes per l'esperimento di fisica delle particelle BTeV", degree
thesis, University "La Sapienza", Rome, 2004 (in Italian). Also
available at
http://www-btev.fnal.gov/cgi-bin/public/DocDB/ShowDocument .

\bibitem
.S. Berardis et al., "Fiber optic sensors for space missions" 2003
IEEE Aerospace Conference Proceeding, Big Sky Montana, March 8-15,
2003, pp. 1661-1668

\bibitem
.L. Benussi et al., "Results of Long-Term Position Monitoring by
Means of Fiber Bragg Grating Sensors for the BTeV Detector",
Frascati preprint LNF - 03 / 15(IR)

\bibitem
.E.Basile et al., "Study of Tensile Response of Kapton, and Mylar
Strips to Ar and CO2 Mixtures for the BTeV Straw Tube Detector",
presented by F. Di Falco at 10th Vienna Conference On
Instrumentation 16-21 Feb 2004, Vienna, Austria, LNF - 04 / 5(P).

\bibitem
.L.Benussi et al., Nucl. Instr. Meth. A361 (1995) 180-191

\bibitem
.A.Balla et al., Nucl. Instr. Meth. A461 (2001) 524-525

\bibitem
.G. Mazzitelli et al., Nucl. Instr. Meth. A515 (2003) 524-542

\end{thebibliography}
\end{document}